\documentclass[journal,10pt]{IEEEtran}

\usepackage[normalem]{ulem}
\usepackage{graphicx}
\usepackage{xcolor}
\usepackage{amssymb}

\usepackage{multirow}

\usepackage{multicol}

\usepackage{diagbox}

\usepackage{tabularx}

\usepackage[cmex10]{amsmath}

\usepackage{url}

\usepackage{dsfont}

\usepackage{pifont}

\usepackage{amsthm}

\usepackage{latexsym}

\usepackage{subfigure}

\usepackage{booktabs}

\usepackage{stfloats}

\usepackage[absolute,showboxes]{textpos}

\TPMargin{3pt}

\newcommand{\copyrightstatement}{
    \begin{textblock}{0.84}(0.08,0.953) 
         \noindent
         \footnotesize
         \copyright 2021 IEEE. Personal use of this material is permitted. Permission from IEEE must be obtained for all other uses, in any current or future media, including reprinting/republishing this material for advertising or promotional purposes, creating new collective works, for resale or redistribution to servers or lists, or reuse of any copyrighted component of this work in other works. DOI: 10.1109/TVT.2021.3129880
    \end{textblock}
}

\begin{document}
\copyrightstatement
\title{
\huge{Joint Power Allocation and Placement Scheme for UAV-assisted IoT with QoS Guarantee}}

%
%
\author{Ruirui~Chen,
        Yanjing~Sun,~\IEEEmembership{Member,~IEEE}, Liping~Liang, and Wenchi~Cheng,~\IEEEmembership{Senior Member,~IEEE}
\thanks{ This work is supported in part by Natural Science Foundation of Jiangsu Province under Grant BK20200650, in part by National Natural Science Foundation of China under Grant 62071472, in part by China Postdoctoral Science Foundation under Grant 2019M660133, in part by Program for ``Industrial IoT and Emergency Collaboration'' Innovative Research Team in CUMT under Grant 2020ZY002, in part by Fundamental Research Funds for the Central Universities under Grant 2019QNB01. (Corresponding author: Yanjing Sun.)
\par R. Chen and Y. Sun are with School of Information and Control Engineering, China University of Mining Technology, Xuzhou, 221116, China (emails: rrchen@cumt.edu.cn, yjsun@cumt.edu.cn). L. Liang and W. Cheng are with State Key Laboratory of Integrated Services Networks, Xidian University, Xi'an, 710071, China (e-mails: liangliping@xidian.edu.cn, wccheng@xidian.edu.cn).
}
}

\markboth{IEEE Transactions on Vehicular Technology, VOL. 71, NO. 1, JANUARY 2022
}%
{Shell \MakeLowercase{\textit{et al.}}: Bare Demo of IEEEtran.cls for IEEE Journals}

\maketitle

\begin{abstract}

In the disaster and remote regions, unmanned aerial vehicles (UAVs) can assist the data acquisition for Internet of Things (IoT). How to cover massive IoT devices (IDs), which require diverse quality-of-service (QoS), is a crucial challenge. For UAV-assisted IoT, this paper studies the deployment scheme with QoS guarantee to place multiple UAVs for covering all ground IDs and maximizing the average data rate of UAVs. First, for the ground ID, we propose the QoS demand based power allocation (QDPA) algorithm to solve the diversity of QoS with respect to data rate demand. Then, the data rate maximization placement (DRMP) algorithm is proposed to optimize the placement of single UAV. Finally, based on QDPA and DRMP algorithms, we propose the joint power allocation and placement (JPAP) scheme with QoS guarantee, which can cover massive IDs, to deploy multiple UAVs for maximizing the average UAV data rate. Simulation experiments are conducted to verify the superiority of our proposed JPAP scheme, which can minimize the UAV number and maximize the average data rate of UAVs.

\end{abstract}


\IEEEpeerreviewmaketitle

\section{Introduction}

\IEEEPARstart{A}{s} important driving force of the fifth generation (5G) communications, Internet of Things (IoT) can achieve the intercommunication among IoT devices (IDs) such as sensors, wearable devices, and intelligent phones \cite{1}. The research in \cite{2} shows that the number of IDs connecting to Internet will be 100 billion by 2025, thus resulting in huge data acquisition of IoT. For traditional networks, terrestrial base stations (BSs) can assist IoT to directly acquire data from IDs. However, due to the inflexible network topology, which is caused by fixed BS deployment and destruction of the communication infrastructure, IoT can hardly achieve the real-time coverage and data acquisition of IDs in the disaster and remote regions.

Unmanned aerial vehicles (UAVs) have been applied into many fields, such as monitoring, surveillance, and cargo delivery \cite{3}. With advantages of mobility, flexibility, and line-of-sight (LoS) channels, UAVs can act as flying BSs to improve the coverage and data acquisition of IoT in the disaster and remote regions, which may be caused by the lack of ground BS \cite{4}. Therefore, to enhance the flexibility and coverage of IoT, UAV-assisted IoT has received much research attention from academia and industry \cite{5}. The authors of \cite{6} studied the resource allocation to optimize the uplink transmission for UAV-assisted IoT. In \cite{7}, the data dissemination performance of UAV-assisted IoT was improved by joint resource allocation and UAV mobility optimization. For UAV-assisted IoT, the maximum energy consumption of IDs was minimized by optimizing the UAV trajectory and time allocation of ID \cite{8}. The authors in \cite{9} proposed the multihop D2D communication and multiantenna transceiver design to satisfy the transmission reliability of UAV-assisted IoT in disaster and hotspot regions. The works of \cite{6,7,8,9} do not consider the scenario where IDs can not be served by single UAV, but the multi-UAV placement that can cover all IDs is urgently required for UAV-assisted IoT.

Recently, UAV deployment schemes, which aim to cover massive IDs, have received much research attention \cite{10,11,12,13,14,15}. For the LoS-dominated ID-UAV channel, multiple UAVs were placed to cover the IDs by using K-means algorithm, and the uncovered IDs were served by ground BSs \cite{10}. To cover all the IDs with minimum UAV number, the spiral deployment scheme was proposed to place the UAVs sequentially along the perimeter of area boundary \cite{11}. By giving high priority to dense boundary ID, the authors in \cite{12} studied the dense boundary prioritized coverage algorithm to place multiple UAVs for the maximization of average UAV capacity.
The works in \cite{10,11,12} assume that all the IDs have the same quality-of-service (QoS) demand. Due to different traffic requirements (e.g., voice call and multimedia transmission), IDs have diverse QoS with respect to data rate demand. Therefore, utilizing the exhaustive search method, the authors in \cite{13} investigated the UAV deployment scheme, which can satisfy the QoS requirement, to maximize the number of covered IDs. In \cite{14}, an on-demand density-driven UAV placement scheme was proposed to maximize the covered ID number under the constraint of minimum QoS requirement. The authors of \cite{15} proposed the utility-aware resource management to maximize the served ID number with different coverage and QoS requirement.

The authors of \cite{13,14,15} ignore the multi-UAV placement scheme with QoS guarantee to cover massive IDs, which can not be served by single UAV. Note that maximizing the average data rate of UAVs can reduce the service time for UAVs. Furthermore, the total transmit power of UAV is limited, and the power allocation is necessary for UAV to improve the energy efficiency. Therefore, this paper studies the efficient multi-UAV coverage scheme with QoS guarantee, which can cover the massive IDs and maximize the average data rate of UAVs. The key contributions of the paper are fourfold, which can be summarized as follows.
\begin{itemize}
  \item The efficient heuristic multi-UAV coverage scheme that can cover all the IDs is proposed from the perspective of the ID's QoS guarantee and UAV data rate.
  \item To solve the diversity of QoS with respect to data rate demand, we propose the QoS demand based power allocation (QDPA) algorithm.
  \item Using QDPA algorithm, the data rate maximization placement (DRMP) algorithm is proposed to obtain the optimal placement for single UAV.
  \item To maximize the average UAV data rate, we propose the joint power allocation and placement (JPAP) scheme, which is based on QDPA and DRMP algorithms, to deploy multiple UAVs for serving all the IDs with diverse QoS requirements.
\end{itemize}

\section{System Model}

In the considered system model, there are $K$ IDs that can be denoted by the set $\mathcal{K}=\{1,2\cdots,K\}$. The $k$-th $(k\in\mathcal{K})$ ID is at the location $(a_k, b_k, 0)$. The UAV set is represented by $\mathcal{N}=\{1,2\cdots,N\}$, and $n$-th ($n\in\mathcal{N}$) UAV is at $(x_n, y_n, H)$. Furthermore, we define $\mathbf{u}_{k}=(a_k, b_k)$ and $\mathbf{v}_{n}=(x_n, y_n)$ as the horizontal coordinates of $k$-th ID and $n$-th UAV, respectively. The IDs are at the ground (i.e., horizontal plane $H=0$) and the UAVs fly at a fixed altitude $H>0$ above the ground. The IDs that are served by the $n$-th UAV are represented by the set $\Omega_n$. The cardinality of set $\Omega_n$ is $K_n=|\Omega_n|$, i.e., the number of IDs that are served by the $n$-th UAV. We assume that $N$ UAVs will be deployed to cover all the $K$ IDs, i.e., $K=\sum\limits_{n\in\mathcal{N}} K_n=\sum\limits_{n\in\mathcal{N}} |\Omega_n|$. The deployment process is managed by the central controller, which can be the mobile BS for the disaster region and the fixed BS for the remote region.


This paper assumes that the ID-UAV channel is LoS-dominated. Thus, the channel gain from the $n$-th UAV to the $k$-th ID can be written as \cite{16,17}
\begin{equation}\label{1}
g_{n,k}=D_{n,k}^{-2}\rho_0=\frac{\rho_0}{\parallel \mathbf{v}_{n}-\mathbf{u}_{k}\parallel^2+H^2},
\end{equation}
where $D_{n,k}$ is the distance from the $n$-th UAV to the $k$-th ID, $\rho_0$ denotes the channel gain of reference distance 1 m and $\|\cdot\|$ is the Euclidean norm of a vector. It is assumed that ID location information is available to UAVs. For the $n$-th UAV, the transmit power allocated to the $k$-th ($k\in\Omega_n$) ID is $P_k$, which follows
\begin{equation}\label{2aa}
\sum_{k\in\Omega_n} P_{k}=P_{\textup{T}},
\end{equation}
where $P_{\textup{T}}$ is the total transmit power of single UAV.

Note that in the disaster and remote regions, the communication requirement is relatively few in general, and thus there are enough bandwidth resources for the UAV-assisted IoT.
Orthogonal bandwidth resources are allocated to the UAVs, which can avoid the interference among UAVs. There may exist the scenario where some IDs are covered by multiple UAVs, but each ID is served by one UAV within its coverage, i.e., $\Omega_{n_1}\cap \Omega_{n_2}=\emptyset$ for $n_1\neq n_2$. To avoid the interference among IDs, the $n$-th UAV provides different resource with equal bandwidth $B$ to each ID in $\Omega_n$. For mathematical simplicity, the bandwidth $B$ is normalized to one.
The achievable data rate of the $k$-th ($k\in\Omega_n$) ID can be expressed as
\begin{equation}\label{2}
R_{n,k}=\textup{log}_2(1+\frac{P_kg_{n,k}}{\sigma^2}),
\end{equation}
where $\sigma^2$ denotes the noise power. The data rate of the $n$-th UAV, which is sum data rate of IDs in $\Omega_n$, can be written as follows:
\begin{equation}\label{2a}
\widehat{R}_n=\sum_{k\in\Omega_n} R_{n,k}.
\end{equation}

The QoS generally refers to the quantitative network measurement, which is a combination of several metrics such as data rate, delay, etc. This paper will study QoS-guaranteed power allocation and UAV placement scheme by providing data rate guarantee for IDs. Due to different traffic requirements, IDs have vastly diverse QoS with respect to data rate demand. For the $k$-th ID, the minimum required data rate is $r_k$. Therefore, to guarantee the QoS of the $k$-th ID (i.e., $r_k$), the maximum UAV coverage radius projected on the ground can be derived as follows:
\begin{equation}\label{2a}
d_k=\sqrt{\frac{P_k\eta}{2^{r_k}-1}-H^2},
\end{equation}
where $\eta=\rho_0/\sigma^2$.

\section{Problem Formulation and Analysis}
This paper denotes the average data rate of UAVs as the ratio of sum UAV data rate to UAV number. The objective of this paper is to maximize the average data rate of UAVs by the power allocation and placement of UAVs. We will determine the deployment of multiple UAVs to serve all the distributed IDs with diverse QoS demands. The formulated problem can be expressed as follows:
\begin{align}
\!\!\!\!\!\!\!\!\!\!\!\!\!\!\!\!\!\!\!\!\!\!\!\!\!\!\!\!\!\!\!\!\!&\boldsymbol{P1}:\mathop{\textup{max}}_{(P_{k}, \mathbf{v}_{n}, N)} ~~\sum_{n\in\mathcal{N}} \frac{\widehat{R}_n}{N}  \tag{6a} \\
&\textup{s.t.}:~~1).~\parallel \mathbf{v}_{n}-\mathbf{u}_{k}\parallel^2\leq d_k^{2}, ~\forall k\in\Omega_n, ~\forall n\in\mathcal{N};  \tag{6b} \label{3}\\
&~~~~~~~~2).~\sum_{k\in\Omega_n} P_{k}=P_{\textup{T}}, ~\forall n\in\mathcal{N};  \tag{6c} \label{6c}\\
&~~~~~~~~3).~\sum\limits_{n\in\mathcal{N}} |\Omega_n|=K;  \tag{6d} \label{4}  \\
&~~~~~~~~4).~|\Omega_n|\leq K_\textup{max}, ~\forall n\in\mathcal{N}.  \tag{6e} \label{6e}
\end{align}  \setcounter{equation}{6}
~The constraint (\ref{3}) denotes that the distances between the IDs and the served UAV can satisfy the QoS of each ID. The constraint (\ref{6c}) represents the total power constraint of single UAV, i.e., the total power of each UAV is $P_{\textup{T}}$. The constraint (\ref{4}) guarantees that the deployed UAVs can serve all the IDs. The constraint (\ref{6e}) means that for each UAV, $K_\textup{max}$ is the maximum number of covered IDs, which is determined by the service ability of UAV.

Problem $\boldsymbol{P1}$ is an NP-hard problem due to the integer optimization variable $N$ (i.e., the UAV number). Moreover, from (\ref{2a}), we can observe that the larger $r_k$ is, the smaller $d_k$ is. The more diverse QoS is, the larger UAV number $N$ will be. The diverse QoS with respect to data rate demand, which results in the difference of $d_k$ corresponding to ID, will significantly complicate the problem $\boldsymbol{P1}$ and make our study nontrivial.
To increase average data rate of UAVs, we should increase UAV data rate $\widehat{R}_n$ and decrease UAV number $N$ by optimizing the UAV placement and maximizing the covered ID number of UAV, respectively.
On this basis, an efficient multi-UAV coverage scheme is proposed with QoS guarantee. For ease of searching, we summarize the notations used throughout the paper in Table I.

%
%
%

\begin{table}  
\normalsize

  \centering
  \caption{\textbf{Summary of Notations} }

  \begin{tabular}{p{1cm} p{6.5cm}}
  \bottomrule

   \hline
$\mathcal{K}$  & Set of IDs   \\
  \hline
$\mathcal{N}$  &  Set of UAVs   \\
  \hline
$\mathbf{u}_{k}$  &  Horizontal coordinates of $k$-th ID  \\
  \hline
$\mathbf{v}_{n}$   &  Horizontal coordinates of $n$-th UAV   \\
\hline
$\Omega_n$   &  Set of IDs that are served by the $n$-th UAV   \\
\hline
$K_n$   &  The number of IDs that are served by the $n$-th UAV   \\
  \hline
$g_{n,k}$   &  Channel gain from the $n$-th UAV to the $k$-th ID   \\
  \hline
$D_{n,k}$   &  Distance from the $n$-th UAV to the $k$-th ID   \\
  \hline
$\rho_0$    &  Channel gain of reference distance 1 m  \\
  \hline
$P_k$   &  Transmit power allocated to the $k$-th ID   \\
  \hline
$P_{\textup{T}}$   &  Total transmit power of UAV   \\
  \hline
$R_{n,k}$   &  Achievable data rate of the $k$-th ID served by $n$-th UAV   \\
  \hline
$\sigma^2$   &  Noise power   \\
  \hline
$\widehat{R}_n$   &  Data rate of the $n$-th UAV   \\
  \hline
$r_k$   &  Minimum required data rate of the $k$-th ID    \\
  \hline
$d_k$   &  Maximum UAV coverage radius projected on the ground   \\
  \hline
$K_\textup{max}$   &  Maximum number of covered IDs for the UAV   \\
  \hline
\bottomrule
\end{tabular}
\end{table}

\section{Joint Power Allocation and Placement Scheme}
In this section, by optimizing the deployment of multiple UAVs, the JPAP scheme is proposed to cover all the IDs and maximize the average data rate of UAVs. Note that the IDs are randomly scattered at the ground and the distances between IDs are different. Therefore, to solve the problem $\boldsymbol{P1}$, we should determine the coverage order of IDs (i.e., cover the IDs in sequence), and then obtain the optimal placement of UAVs. The constraints (\ref{4}) and (\ref{6e}) of problem $\boldsymbol{P1}$ are satisfied by the coverage order of IDs. In other words, the determined coverage sequence of IDs can guarantee the constraints (\ref{4}) and (\ref{6e}) of problem $\boldsymbol{P1}$. Then, we can decompose the problem $\boldsymbol{P1}$ into $N$ optimization subproblems of UAV placement, which can maximize the data rate of UAV.
To solve the subproblem, the QDPA algorithm is first given to solve the diversity of QoS with respect to data rate demand, and then we derive the DRMP algorithm to obtain the optimal placement of single UAV.

We will deploy multiple UAVs sequentially based on the coverage order of IDs, which can be explained as follows. The boundary IDs are defined as the ID with maximum $a_k$, ID with maximum $b_k$, ID with minimum $a_k$, and ID with minimum $b_k$. Obviously, we allocate the total power $P_{\textup{T}}$ to the first covered ID, and thus the coverage radius is $d_k^0=\sqrt{(P_{\textup{T}}\eta)/(2^{r_k}-1)-H^2}$. Furthermore, the first covered ID chosen from the boundary IDs is the dense boundary ID, which has the most IDs within its $2d_k^0$ area. The high priority is given to cover the dense boundary ID, which will decrease the occurrence probability of outlier IDs that need one dedicated UAV for its coverage.
For the $n$-th UAV, we first search for the dense boundary ID $k_0$ of the uncovered ID set $\mathcal{K}_{\textup{U}}$ and the IDs within $2d_{k_0}^0$ area of $k_0$, which is represented by the ID set $\mathcal{K}_0$. Then, the $n$-th UAV covers the dense boundary ID $k_0$. Finally, to cover as many uncovered IDs in $\mathcal{K}_0$ as possible, QDPA and DRMP algorithms are proposed to place the $n$-th UAV inwards toward the area center of uncovered ID set $\mathcal{K}_{\textup{U}}$. After placing the $n$-th UAV, the area of uncovered ID set $\mathcal{K}_{\textup{U}}$ will shrink at the local area near $k_0$. The above process is repeated until all the IDs are covered, i.e., $\mathcal{K}_{\textup{U}}=\emptyset$.

Then, for the $n$-th UAV, the data rate maximization subproblem can be written as
\begin{align}
\!\!\!\!\!\!\!\!\!\!\!\!\!\!\!\!\!\!\!\!\!\!\!\!\!\!\!\!\!\!\boldsymbol{P2}&:\mathop{\textup{max}}_{(P_{k}, x_{n}, y_{n})} ~~\widehat{R}_n=\sum_{k\in\Omega_n} R_{n,k}  \tag{7a} \\     
\textup{s.t.}:~~&1).~(x_n-a_k)^2+(y_n-b_k)^2\leq d_k^{2}, ~\forall k\in\Omega_n;   \tag{7b} \label{5} \\
&2).~\sum_{k\in\Omega_n} P_{k}=P_{\textup{T}}.  \tag{7c} \label{3a}
\end{align} \setcounter{equation}{7}
To maximize average data rate of UAVs, we propose the JPAP scheme to determine the coverage of IDs, and then obtain the optimal placement of single UAV by using the proposed QDPA and DRMP algorithms, which are given in the following.
\subsection{QoS Demand Based Power Allocation}
To cope with QoS diversity, we employ the adaptive power allocation to make the $d_k$ of different IDs have the same value $d_{\textup{sv}}$, which can be expressed as follows:
\begin{equation}\label{4a}
d_{\textup{sv}}=d_1=d_2=\cdots=d_k.
\end{equation}
Then, we can derive
\begin{equation}\label{4aa}
\frac{2^{r_k}-1}{2^{r_1}-1}P_1=\frac{2^{r_k}-1}{2^{r_2}-1}P_2=\cdots=P_k.
\end{equation}
Based on (\ref{3a}), the QDPA algorithm can be written as
\begin{equation}\label{4aaa}
P_k=\frac{P_{\textup{T}}(2^{r_k}-1)}{\sum\limits_{k\in\Omega_n} (2^{r_k}-1)}.
\end{equation}
Utilizing the QDPA algorithm, i.e., $P_k$ in (\ref{4aaa}), we can obtain
\begin{equation}\label{4aaaa}
d_{\textup{sv}}=\sqrt{\frac{P_{\textup{T}}\eta}{\sum\limits_{k\in\Omega_n} (2^{r_k}-1)}-H^2}.
\end{equation}
Thus, to meet the high QoS demand of ID, UAV will allocate large power to the ID with high QoS demand. Problem $\boldsymbol{P2}$ can be transformed into the following problem:
\begin{align}
\!\!\!\!\!\!\!\!\!\!\!\!\!\!\!\!\!\!\!\!\!\!\!\!\!\!\!\!\!\!&\boldsymbol{P3}:\mathop{\textup{max}}_{(x_{n}, y_{n})} ~~\widehat{R}_n=\sum_{k\in\Omega_n} R_{n,k}  \tag{12a} \\     
\textup{s.t.}&:~~(x_n-a_k)^2+(y_n-b_k)^2\leq d_{\textup{sv}}^{2}, ~\forall k\in\Omega_n.   \tag{12b} \label{5} 
\end{align} \setcounter{equation}{12}
~Since the problem $\boldsymbol{P3}$ is not concave, we are interested in the function $\textup{log}_2(1 + z)$ where $z\geq 0$, the tight lower-bound of which is function $\tau\textup{log}_2z+\beta$ \cite{18}.
Furthermore, the coefficients $\tau$ and $\beta$ are chosen as $\tau=\frac{z_0}{1+z_0}$ and $\beta=\textup{log}_2(1+z_0)-\frac{z_0}{1+z_0}\textup{log}_2z_0$.
Since $z_0\geq0$, $\tau$ is not less than 0. The lower-bound $\tau\textup{log}_2z+\beta$ is equal to $\textup{log}_2(1 + z)$ for $z=z_0$.
The function $R_{n,k}$ can be relaxed as
\begin{equation}\label{10}
\widetilde{R}_{n,k}=\tau_k\textup{log}_2P_k\eta+\beta_k-\tau_k J_{n,k},
\end{equation}
where $J_{n,k}=\textup{log}_2[(x_n-a_k)^2+(y_n-b_k)^2+H^2]$, $\tau_k$ and $\beta_k$ are the coefficients corresponding to the $k$-th ID.

\subsection{DRMP Algorithm for Single UAV}
Based on (\ref{10}), the problem $\boldsymbol{P3}$ can be converted into
\begin{equation}\label{11}
\boldsymbol{P4}:\mathop{\textup{min}}_{(x_{n}, y_{n})} ~~\sum_{k\in\Omega_n} \tau_k J_{n,k}
\end{equation}
subject to the constraint (\ref{5}). The DRMP algorithm for single UAV is proposed to solve problem $\boldsymbol{P4}$ by maximizing the UAV data rate.

The Hessian matrix of $J_{n,k}$ is expressed as follows:
\begin{equation}\label{12}
 \boldsymbol{\textup{T}_j}=\left[
                                        \begin{array}{cc}
                                          t_{j1} & t_{j2} \\
                                          t_{j3} & t_{j4} \\
                                        \end{array}
                                      \right]=\left[
                                                \begin{array}{cc}
                                                  \frac{\partial^2J_{n,k}}{\partial x_n^2} & \frac{\partial^2J_{n,k}}{\partial x_n\partial y_n} \\
                                         \frac{\partial^2J_{n,k}}{\partial y_n \partial x_n} & \frac{\partial^2J_{n,k}}{\partial y_n^2}  \\
                                                \end{array}
                                              \right].
\end{equation}
We can derive $t_{j1}$ and $t_{j1}t_{j4}-t_{j2}t_{j3}$ as
\begin{equation}\label{13}
  t_{j1}=\frac{2\left[-(x_n-a_k)^2+(y_n-b_k)^2+H^2\right]}{\textup{ln}2D_{n,k}^{4}}
\end{equation}
and
\begin{equation}\label{14}
\begin{aligned}
  t_{j1}t_{j4}-t_{j2}t_{j3}=\frac{4\left\{-\left[(x_n-a_k)^2+(y_n-b_k)^2\right]^2+H^4\right\}}{\left(\textup{ln}2D_{n,k}^{4}\right)^2}.
\end{aligned}
\end{equation}
From (\ref{13}) and (\ref{14}), we can see that problem $\boldsymbol{P4}$ will be convex when $(x_n-a_k)^2+(y_n-b_k)^2<H^2$ holds i.e., $d_{\textup{sv}}$ is less than $H$.

For the case $d_{\textup{sv}}<H$, we can formulate the Lagrange function of $\boldsymbol{P4}$ as follows:
\begin{equation}\label{15}
\begin{aligned}
\!\!\!\!\!L(x_n,&y_n;\boldsymbol{\lambda})=\sum_{k\in\Omega_n}\tau_k J_{n,k}\\
&+\sum_{k\in\Omega_n}\lambda_k \left[(x_n-a_k)^2+(y_n-b_k)^2-d_{\textup{sv}}^2\right],
\end{aligned}
\end{equation}
where $\boldsymbol{\lambda}$ denotes the Lagrange multiplier vector, which consists of $\lambda_k$ ($\forall k\in\Omega_n$). Since problem $\boldsymbol{P4}$ is a strictly convex optimization problem, $\boldsymbol{P4}$'s optimal Lagrange multiplier vector $\boldsymbol{\lambda}^*$ is also the optimal solution to $\boldsymbol{P4}$'s dual problem $\boldsymbol{P4D}$, which can be expressed as follows:
\begin{align}
\!\!\!\!\!\!\boldsymbol{P4D}&:\mathop{\textup{max}}_{\boldsymbol{\lambda}} ~~\Big\{\widetilde{L}(\boldsymbol{\lambda})=\mathop{\textup{min}}_{(x_n,y_n)} ~~L(x_n,y_n;\boldsymbol{\lambda})\Big\}  \tag{19a} \label{16a}\\     
\textup{s.t.}&:~~\lambda_k\geq 0, ~\forall k\in\Omega_n,   \tag{19b} \label{16}
\end{align}  \setcounter{equation}{19}
\!\!where $\widetilde{L}(\boldsymbol{\lambda})$ denotes the Lagrange dual function. Based on convex optimization theory, function $\widetilde{L}(\boldsymbol{\lambda})$ is concave on the space spanned by $\boldsymbol{\lambda}$. Thus, we use the subgradient method to track $\boldsymbol{\lambda}^*$, which is given as follows:
\begin{equation}\label{18}
\lambda_k^*=\left\{\lambda_k^*+\epsilon_\lambda\left[(x_n-a_k)^2+(y_n-b_k)^2-d_{\textup{sv}}^2\right]\right\}^+,
\end{equation}
where $\epsilon_\lambda$ is positive real-valued number arbitrarily close to 0 and $\{\lambda\}^+$ denotes the maximum value between $\lambda$ and 0.
The iteration of (\ref{18}) converges to $\boldsymbol{\lambda}^*$. Then, by taking $\boldsymbol{\lambda}^*$ into
\begin{equation}\label{19}
\frac{\partial L(x_n,y_n;\boldsymbol{\lambda})}{\partial x_n}\bigg|_{\boldsymbol{\lambda}=\boldsymbol{\lambda}^*}=0
\end{equation}
and
\begin{equation}\label{20}
\frac{\partial L(x_n,y_n;\boldsymbol{\lambda})}{\partial y_n}\bigg|_{\boldsymbol{\lambda}=\boldsymbol{\lambda}^*}=0,
\end{equation}
we obtain the optimal placement $\mathbf{v}_{n}^*$. Using $\mathbf{v}_{n}^*=(x_n^*,y_n^*)$, we can derive the maximum UAV data rate $\widehat{R}_n^*$.

For the general case $d_{\textup{sv}}\geq H$, the feasible region can be divided into inner feasible region $\mathcal{I}=\{(x_n,y_n)|(x_n-a_k)^2+(y_n-b_k)^2< H^2, ~\forall k\in\Omega_n\}$ and outer feasible region $\mathcal{O}=\{(x_n,y_n)|H^2\leq(x_n-a_k)^2+(y_n-b_k)^2\leq d_{\textup{sv}}^2, ~\forall k\in\Omega_n\}$. For $\mathcal{I}$, we first use Lagrange method to derive the optimal placement $\mathbf{v}_{n}^{\textup{~i}}$ and maximum UAV data rate $\widehat{R}_n^{\textup{~i}}$. Then, for $\mathcal{O}$, the exhaustive search method is used to obtain the optimal placement $\mathbf{v}_{n}^{\textup{o}}$ and maximum UAV data rate $\widehat{R}_n^{\textup{o}}$. Finally, the larger one between $\widehat{R}_n^{\textup{~i}}$ and $\widehat{R}_n^{\textup{o}}$ is the maximum UAV data rate $\widehat{R}_n^*$, and the optimal placement corresponding to $\widehat{R}_n^*$ is the optimal placement of single UAV $\mathbf{v}_{n}^*$.

Based on the above analysis, we propose the DRMP algorithm, which is summarized as follows: 1) for $d_{\textup{sv}}< H$, the Lagrange method is used to calculate the optimal placement and maximum UAV data rate; 2) for $d_{\textup{sv}}\geq H$, we utilize the Lagrange method and exhaustive search method to obtain the optimal placement and maximum UAV data rate. Note that $\mathcal{I}=\mathcal{O}=\emptyset$ indicates that the feasible region does not exist, i.e., the $n$-th UAV can not cover all the $K_n$ IDs.
\subsection{Illustration of JPAP scheme}
To maximize average data rate of UAVs, we propose the JPAP scheme, which covers all the $K$ IDs, to solve problem $\boldsymbol{P1}$ based on QDPA and DRMP algorithms. The proposed JPAP scheme can be summarized as follows:
\par \noindent\rule[0.25\baselineskip]{8.8cm}{0.5pt}
\uline{\textbf{Joint power allocation and placement Scheme}}
\par \noindent\emph{\textbf{Initialization}}:
\par \noindent~1) UAV set $\mathcal{N}=\emptyset$ and $n=1$;
\par \noindent~2) ID set $\mathcal{K}$ and ID location $\{\mathbf{u}_{k}\}_{k\in\mathcal{K}}$;
\par \noindent~3) Uncovered ID set $\mathcal{K}_{\textup{U}}=\mathcal{K}$ and covered ID set $\mathcal{K}_{\textup{C}}=\emptyset$;
\par \noindent\emph{\textbf{Iteration}}:
\par \noindent~1) \textbf{while} $\mathcal{K}_{\textup{U}}\neq\emptyset$ \textbf{do}
\par \noindent~2) ~~For $\mathcal{K}_{\textup{U}}$, search for $k_0$ (dense boundary ID) and $\mathcal{K}_0$ (ID \par\noindent~~~~~~set within $k_0$'s $2d_{k_0}^0$ area);
\par \noindent~3) ~~Set $\mathbf{v}_{n}^*=\mathbf{u}_{k_0}$, $\widehat{R}_{n}^*=\textup{log}_2(1+\frac{P_{\textup{T}}\eta}{H^2})$, $\mathcal{K}_{\textup{C}}=k_0$ and $i=1$;
\par \noindent~4) ~~\textbf{while} ($\mathcal{K}_0\neq\emptyset~\&~i\leq K_\textup{max}$) \textbf{do}
\par \noindent~5) ~~~~Search for ID $k_1\in\mathcal{K}_0$ with shortest distance to $\mathbf{v}_{n}^*$;
\par \noindent~6) ~~~~Set $\mathcal{K}_{\textup{C}}=\mathcal{K}_{\textup{C}}\cup k_1$, $\mathcal{K}_0=\mathcal{K}_0 \backslash\ k_1$ and $i=|\mathcal{K}_{\textup{C}}|$;
\par \noindent~7) ~~~~For \,$\mathcal{K}_{\textup{C}}$\,, based\, on\, QDPA\,, calculate\, power allocation\par\noindent~~~~~~~~~$P_k$\,\, and\,\, $d_{\textup{sv}}$\,\, by\,\, using\,\, (\ref{4aaa})\,\, and\,\, (\ref{4aaaa})\,\,, respectively\,;\par\noindent~~~~~~~~\,obtain\,\, inner\,\, feasible\,\, region \,\,$\mathcal{I}$\,\, and\,\, outer \,\,feasible\par\noindent~~~~~~~~\,region $\mathcal{O}$ for $\mathcal{K}_{\textup{C}}$;
\par \noindent~8) ~~~~\textbf{if}   $\mathcal{I}=\mathcal{O}=\emptyset$
\par \noindent~9) ~~~~~~Set $\mathcal{K}_{\textup{C}}=\mathcal{K}_{\textup{C}}\backslash k_1$, \textbf{go to 14)};
\par \noindent10) ~~~~\textbf{else}
\par \noindent11) ~~~~~~To \,\,cover \,\,$\mathcal{K}_{\textup{C}}$\,, \,use\,\, DRMP\,\, algorithm\,\,\, to\,\,\, update  \par\noindent~~~~~~~~~~~optimal placement $\mathbf{v}_{n}^*$ and maximum UAV data rate  \par\noindent~~~~~~~~~~\,\,$\widehat{R}_{n}^*=\sum\limits_{k\in\mathcal{K}_{\textup{C}}} \widetilde{R}_{n,k}$;
\par \noindent12) ~~~~\textbf{end if}
\par \noindent13) ~~\textbf{end while}
\par \noindent14) ~~Set $\mathcal{K}_{\textup{U}}=\mathcal{K}_{\textup{U}}\backslash \mathcal{K}_{\textup{C}}$, $\mathcal{N}=\mathcal{N}\cup n$ and $n=n+1$;
\par \noindent15) \textbf{end while}
\par \noindent16) Obtain optimal UAV placement $\{\mathbf{v}_{n}^*\}_{n\in\mathcal{N}}$ and calculate\par\noindent~~~~ average data rate $\sum\limits_{n\in\mathcal{N}} \frac{\widehat{R}_n^*}{|\mathcal{N}|}$.
\par \noindent\rule[0.25\baselineskip]{8.8cm}{1pt}
\subsection{Complexity Analysis of JPAP Scheme}
For the JPAP scheme, the complexity of the iteration loop is bounded by O($\frac{K(K+1)}{2}$). The subgradient method, which is used to update Lagrange multiplier vector $\boldsymbol{\lambda}$, converges to the desired condition only after O($\frac{1}{\epsilon^2}$) iterations, where $\epsilon$ is the maximum error tolerance \cite{19}. The complexity of the exhaustive search method can be expressed as O($\frac{1}{\epsilon}$). Thus, the complexity of each iteration is bounded by O($\frac{1}{\epsilon^2}$), which is determined by the higher complexity between the Lagrange method and exhaustive search method. In summary, the complexity of the JPAP scheme can be bounded by O($\frac{K(K+1)}{2\epsilon^2}$).

\section{Simulation Results}
Simulation results are presented to evaluate the performance of proposed JPAP scheme. The QoS demand $r_k$ is uniformly distributed in the region $[15,20]$. We set the total transmit power, noise power and channel gain of the reference distance $1~\textup{m}$ as $P_{\textup{T}}=40~\textup{dBm}$, $\sigma^2=-90~\textup{dBm}$ and $\rho_0=-30~\textup{dB}$, respectively. The IDs are uniformly scattered in $1500~\textup{m}\times1500~\textup{m}$ area. The UAV altitude $H$ is $50~\textup{m}$. Note that the proposed JPAP scheme is based on QDPA algorithm in this paper. The illustration of JPAP scheme is given in Fig. 2, which describes the deployment of $N$ UAVs and the placement of first UAV. For comparison, we also give the equal power allocation (EPA) algorithm based JPAP scheme, in which the total transmit power of UAV is equally allocated to covered IDs.

\begin{figure}[thp]
\setlength{\abovecaptionskip}{0.cm}
\setlength{\belowcaptionskip}{-0.cm}
\centering
\vspace{-0.25cm}
\setlength{\abovecaptionskip}{0.cm}
\setlength{\belowcaptionskip}{-0.cm}
\includegraphics[height=2.6in,width=8.2cm]{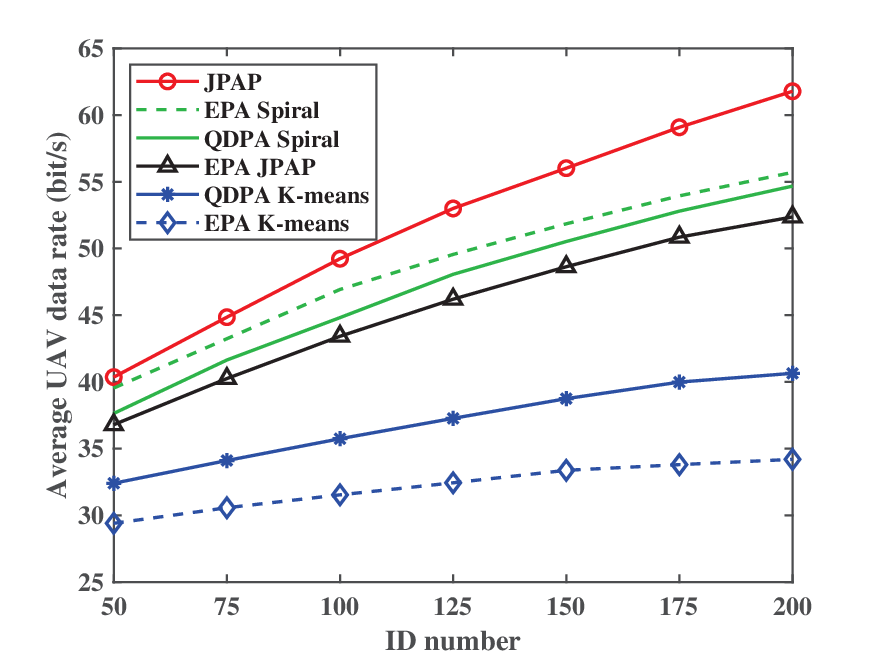}
\caption{\!Average UAV data rate versus ID number}
\label{fig:a}
\intextsep=1pt plus 3pt minus 1pt
\end{figure}

Figure 1 compares the proposed JPAP scheme with EPA algorithm based JPAP scheme, QDPA/EPA algorithm based K-means scheme and QDPA/EPA algorithm based spiral scheme. From Fig. 1,  we can observe that the proposed JPAP scheme and spiral scheme achieve larger average UAV data rates than the K-means scheme. This is because UAVs are placed flexibly based on ID location information. The proposed JPAP scheme obtains the largest average UAV data rate, which benefits from QDPA and DRMP algorithms for each UAV.

\begin{figure}[thp]
\setlength{\abovecaptionskip}{0.cm}
\setlength{\belowcaptionskip}{-0.cm}
\centering
\vspace{-0.35cm}
\setlength{\abovecaptionskip}{0.cm}
\setlength{\belowcaptionskip}{-0.cm}
\includegraphics[height=2.96in,width=8.2cm]{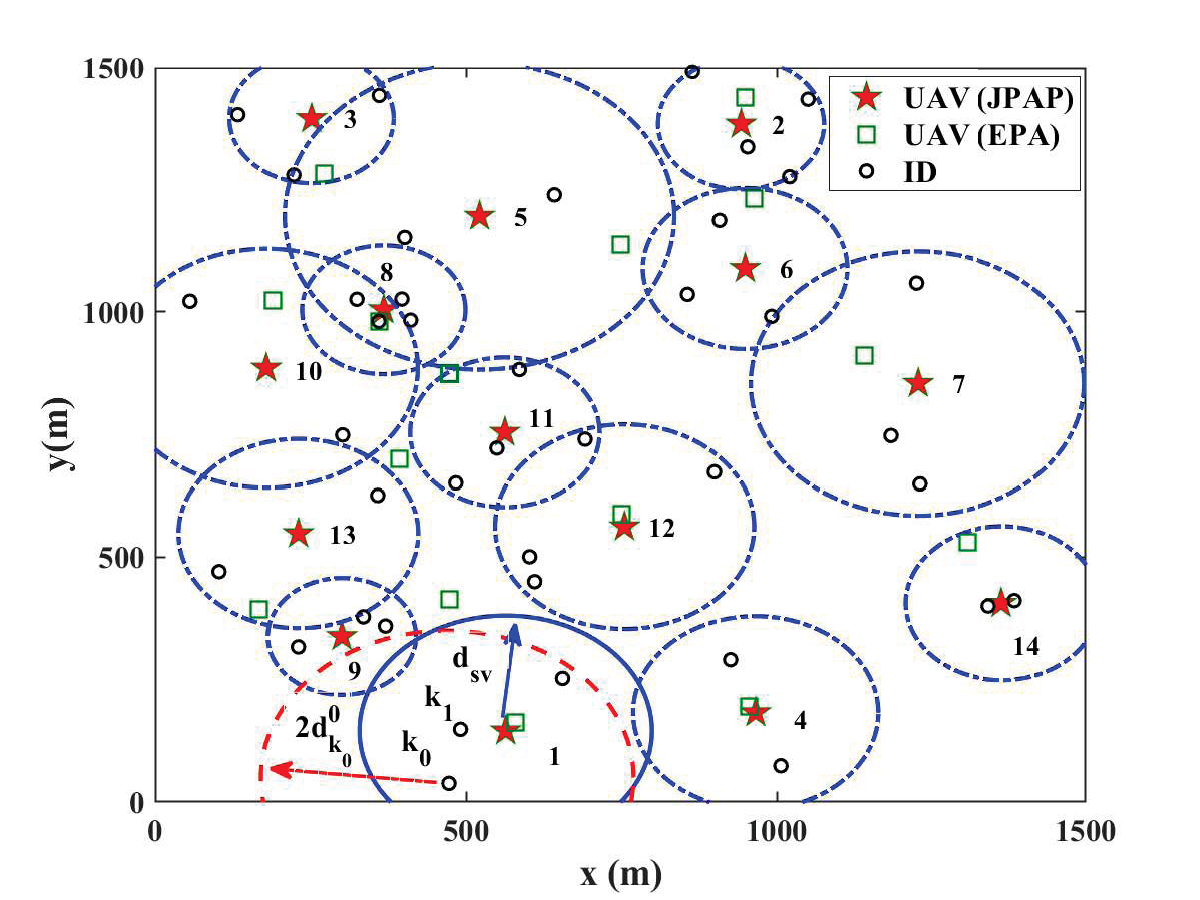}
\caption{\!Demonstration of proposed JPAP scheme and EPA algorithm based JPAP scheme}
\label{fig:a}
\intextsep=1pt plus 3pt minus 1pt
\end{figure}

In Fig. 2, we use the proposed JPAP scheme and EPA algorithm based JPAP scheme to deploy multiple UAVs for covering 40 IDs.
The proposed JPAP scheme requires 14 UAVs denoted by \ding{73}, and the blue dashed circle with radius $d_{\textup{sv}}$ is the coverage of UAV, which is different for different UAVs. The EPA algorithm based JPAP scheme requires 15 UAVs marked by $\Box$, which is larger than the UAV number of proposed JPAP scheme. This is because the proposed JPAP scheme utilizes the adaptive power allocation based on QoS demand.

\begin{figure}[thp]
\setlength{\abovecaptionskip}{0.cm}
\setlength{\belowcaptionskip}{-0.cm}
\centering
\vspace{-0.25cm}
\setlength{\abovecaptionskip}{0.cm}
\setlength{\belowcaptionskip}{-0.cm}
\includegraphics[height=2.6in,width=8.2cm]{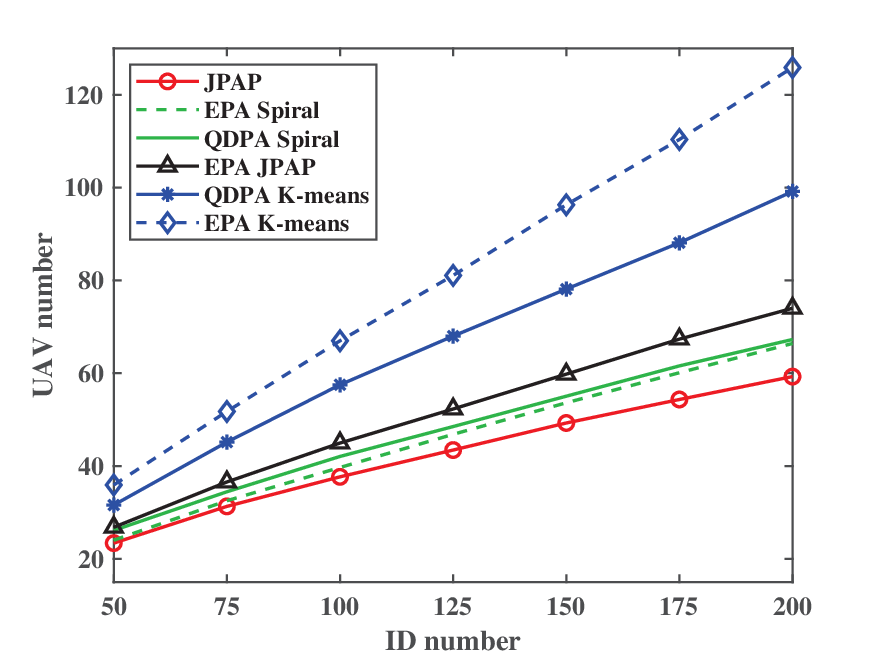}
\caption{\!UAV number under various ID number}
\label{fig:a}
\intextsep=1pt plus 3pt minus 1pt
\end{figure}

 Figure 3 depicts the UAV number curves of the proposed JPAP scheme with EPA algorithm based JPAP scheme, QDPA/EPA algorithm based K-means scheme and QDPA/EPA algorithm based spiral scheme. It can be seen that proposed JPAP scheme obtains the smallest UAV number. Furthermore, the UAV numbers of proposed JPAP scheme and spiral scheme are both less than that of K-means scheme. This is due to the flexible placement of UAV according to ID location.
\section{Conclusions}

The deployment scheme of UAV, which can satisfy the QoS demand of ID, was investigated for UAV-assisted IoT to cover all the IDs and maximize the average data rate of UAVs. To solve the QoS diversity of IDs, the QDPA algorithm was derived based on the adaptive power allocation. We presented the DRMP algorithm to obtain the optimal placement of single UAV. Then, jointly utilizing QDPA and DRMP algorithms, we proposed the efficient multi-UAV coverage scheme, which can maximize the average UAV data rate, to cover all the IDs with diverse QoS demands. Simulation results were provided to validate that proposed JPAP scheme can obtain the minimum UAV number and maximum average data rate of UAVs.

\end{document}